\documentclass{iaus}
\usepackage{graphicx}

\title[AS 12.~~Star Formation Relation] 
{The Star Formation Relation \\ in Nearby Galaxies}

\author[Andreas Schruba]   
{Andreas Schruba$^1$}

\affiliation{$^1$Cahill Center for Astronomy and Astrophysics, California Institute of Technology, \\ MC 249-17, 1200 E California Blvd, Pasadena, CA 91125, USA \\ email: {\tt schruba@astro.caltech.edu}}

\pubyear{2012}
\volume{292}  
\pagerange{1--8}
\jname{Molecular Gas, Dust, and Star Formation in Galaxies}
\editors{Tony Wong \& Juergen Ott, eds.}
\begin{document}

\maketitle

\begin{abstract}
I review observational studies of the large-scale star formation process in nearby galaxies. A wealth of new multi-wavelength data provide an unprecedented view on the interplay of the interstellar medium and (young) stellar populations on a few hundred parsec scale in 100+ galaxies of all types. These observations enable us to relate detailed studies of star formation in the Milky Way to the zoo of galaxies in the distant universe. Within the disks of spiral galaxies, recent star formation strongly scales with the local amount of molecular gas (as traced by CO) with a molecular gas depletion time of $\sim$2 Gyr. This is consistent with the picture that stars form in giant molecular clouds that have about universal properties. Galaxy centers and star-bursting galaxies deviate from this normal trend as they show enhanced star formation per unit gas mass suggesting systematic changes in the molecular gas properties and especially the dense gas fraction. In the outer disks of spirals and in dwarf galaxies, the decreasing availability of atomic gas inevitably limits the amount of star formation, though with large local variations. The critical step for the gas-stars circle seems therefore the formation of a molecular gas phase that shows complex dependencies on various environmental properties and are nowadays investigated by intensive simulational work.
\end{abstract}

\firstsection
\section{Introduction to the Star Formation Relation}

Knowledge of the relationship between star formation and gas mass in galaxies is of great importance to our understanding of galaxy evolution. Over the past 50 years numerous observational and theoretical studies have been devoted to analyze and constrain this relationship. Detailed studies of the gas-stars interaction in our own Galaxy have been feasible over a long time by now and delivered tremendous insight into the various processes and couplings on small spatial scales (for recent reviews see \cite{McKee07} and \cite{Kennicutt12}). Progressive technological developments have now also opened up observational research of the star formation process outside our Galaxy ranging from detailed studies of nearby galaxies to a rapidly increasing number of distant galaxies. With the expectancy of further developments of receiver arrays and sensitive high resolution instruments as ALMA, CCAT, ELT, GMT, and JWST this science field is looking forward to a continued rapid evolution in the near future.

The original proposal of a direct relationship between star formation rate (SFR) and gas mass has been brought forward by \cite{Schmidt59} more than 50 years ago who suggested a power law relationship between the volume densities of SFR and gas mass. However, because volume densities are difficult to constrain observationally in external galaxies, most studies by today have analyzed the power law relationship between surface density quantities. It has been frequently argued that the two relationships will translate to another under the common (though questionably) assumption of constant disk scale heights. However, this overlooks that the two relations operate over vastly different spatial scales. The relation between volume densities extends from the low density diffuse interstellar medium (ISM) to individual high density cores of giant molecular clouds (GMCs) where stars actually form. The relation between surface densities operates on significantly larger scales, namely the average over a wide range of local conditions extending up to the whole galaxy. As the processes that govern the evolution of the ISM and star formation are highly non-linear, it can neither be expected that the two relations are equal nor that one relation can be inferred from the other. It therefore requires simulations to identify the governing physical processes from the observed surface density relationships -- a task that has become feasible within the last years.

The first comprehensive observational studies of such a power law relationship have been performed by \cite{Kennicutt89} and \cite{Kennicutt98} which revealed a tight scaling between the SFR surface density, $\Sigma_{\rm SFR}$, and the total gas surface density, $\Sigma_{\rm Gas} = \Sigma_{\rm HI} + \Sigma_{\rm H2}$. For galaxy-averaged quantities Kennicutt measured a power law relationship with slope $N \approx 1.4$ extending over more than 5 orders of magnitude. The non-linear nature of the relation meant that the star formation efficiency, SFE = $\Sigma_{\rm SFR}/\Sigma_{\rm Gas}$, or its inverse, the total gas depletion time, $\tau_{\rm Dep} =\Sigma_{\rm Gas}/\Sigma_{\rm SFR}$, is not constant but SFR precedes more rapidly in regions of high gas (surface) density. A slope of $N = 1.5$ naturally arises if large-scale gravitational disk instabilities define the timescale of star formation. However, a similar super-linear scaling can arise if dynamical timescales are considered.

\section{The Resolved Linear Molecular Star Formation Relation}

\begin{figure}[t]
\begin{center}
\includegraphics[width=\textwidth]{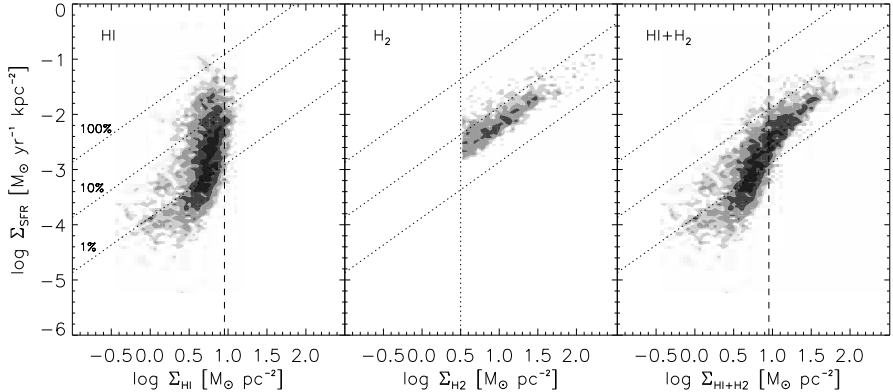} 
\caption{The star formation relation for different gas types: HI (left), H$_2$ (middle),  and HI+H$_2$ (right) for 750 pc sized pixels in 7 nearby spiral galaxies. The contours represent data density. The relationship for different gas types is fundamentally different: HI saturates at $\sim$10 M$_\odot$~pc$^{-2}$ and does not correlate with SFR except in the far outer disks of decreasing HI surface density. H$_2$ scales roughly linear with SFR with a depletion time of $\sim$2 Gyr. The relationship for total gas (HI+H$_2$) is the superposition of the two distinct scalings and exhibits a sharp turnover between HI- and H$_2$-dominated regimes (dashed line). Figure taken from \cite{Bigiel08}.}
\label{fig1}
\end{center}
\end{figure}

While disk-averaged studies have the potential to reveal differences in the global ratio of gas mass and SFR as function of galaxy wide properties, they have limited ability to constrain the underlying (local) physical processes. Today a wealth of new multi-wavelength data provide us an unprecedented detailed view on the star formation process on scales of a few hundred parsec across more than 100 nearby galaxies of all types. Especially the combined analysis of large homogeneous data sets allow to overcome previous large methodological discrepancies between different studies. One such initiative combines the observational data from THINGS (HI; \cite{Walter08}), HERACLES (CO; \cite{Leroy09}), SINGS \& LVL (IR; \cite{Kennicutt03}; \cite{Dale09}), and GALEX (UV; \cite{GildePaz07}). In a first analysis \cite{Bigiel08} combined these data to compare HI, H$_2$, and SFR on 750 pc resolution in a sample of 7 nearby spiral galaxies. The resulting $\Sigma_{\rm SFR} - \Sigma_{\rm Gas}$ relation (Fig.\,\ref{fig1}) consists of two quite distinct regimes. The relation between $\Sigma_{\rm SFR}$ and $\Sigma_{\rm H2}$ is observed to be about linear ($N = 1.0 \pm 0.1$) with mean molecular gas depletion time $\tau_{\rm Dep} = 2.0$ Gyr. The relation between $\Sigma_{\rm SFR}$ and $\Sigma_{\rm HI}$ however is very steep with $\Sigma_{\rm HI}$ saturating at $\sim$10 M$_\odot$~pc$^{-2}$ and $\Sigma_{\rm SFR}$ varying by more than 2 orders of magnitude over a small range of $\Sigma_{\rm HI}$. This suggests that the SFR per unit H$_2$ gas (i.e. SFE) is constant throughout the environments present in the disks of spiral galaxy. \cite{Leroy08} tested if SFE shows dependencies on ISM pressure, dynamical time, galactocentric radius, stellar and gas mass but found none. On the other hand, the ratio $\Sigma_{\rm H2}/\Sigma_{\rm HI}$ is found to correlat with many of these parameters. This suggests that the formation of stars from H$_2$ (inside of GMCs) is largely independent of environment while the formation of H$_2$ out of HI depends strongly on environment.

\begin{figure}[t]
\begin{center}
\begin{minipage}[b]{0.4\textwidth}
\includegraphics[width=1\textwidth]{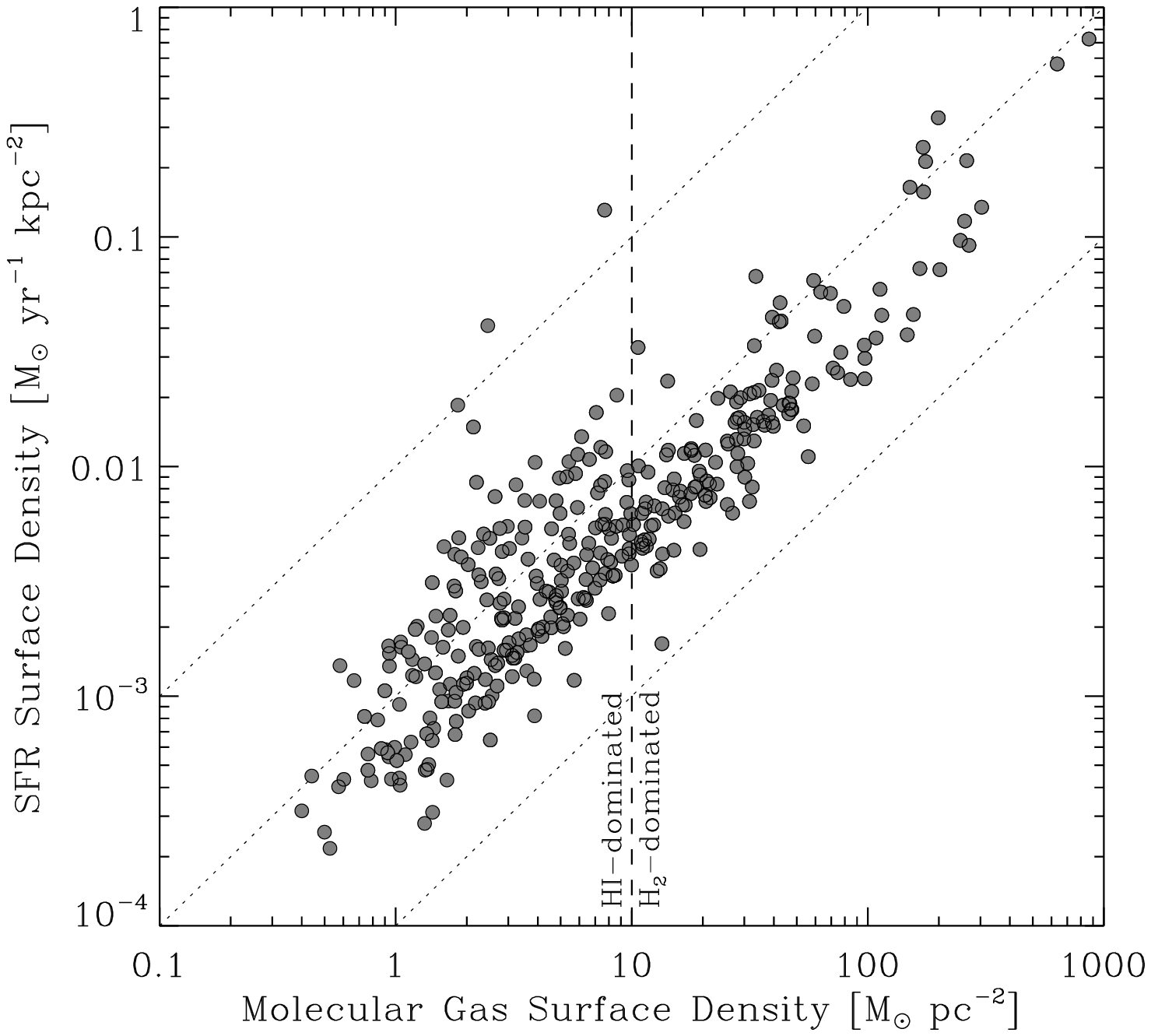}
\end{minipage}
\begin{minipage}[b]{0.4\textwidth}
\includegraphics[width=1\textwidth]{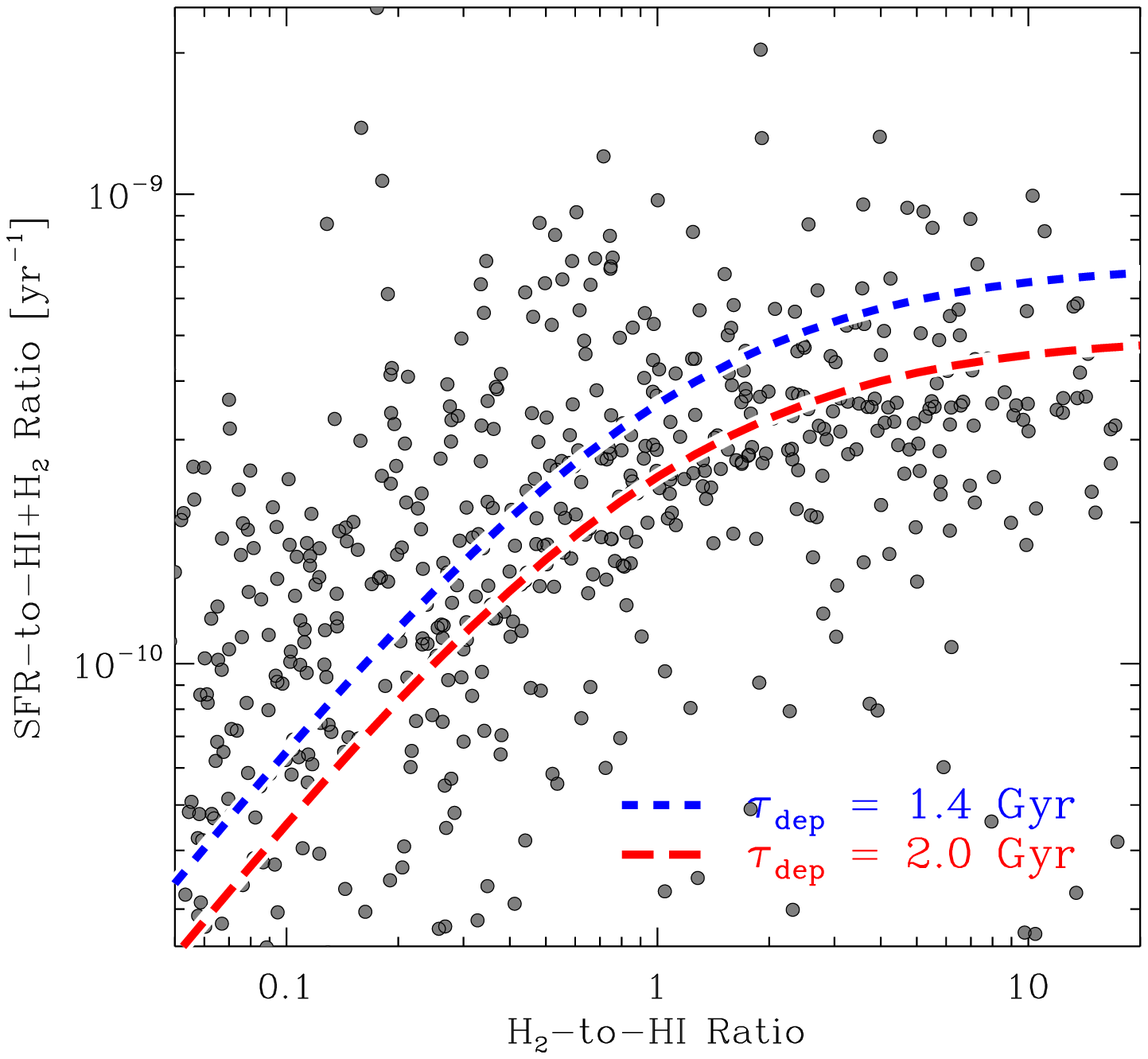}
\end{minipage}
\caption{Left: The molecular star formation relation for a sample of 30 nearby spiral galaxies. Using sensitive CO data from HERACLES and a novel technique to stack spectra that allows us to measure faint CO emission with high significance in regions dominated by HI, we find that SFR and H$_2$ is linearly correlated not only in regions there the ISM is predominately molecular but also in regions where most of the ISM is atomic highlighting the critical role of the molecular gas phase as a prerequisite for star formation. Right: The star formation efficiency -- the ratio of SFR and total gas mass (HI+H$_2$) -- as function of the molecular-atomic phase ratio (H$_2$/HI). The turnover eminent in the total gas-SFR relation is fundamentally related to a smooth transition of the ISM being H$_2$- or HI-dominated. The dashed lines show the expected scaling if SFR depends exclusively and linearly on H$_2$ mass. Figures taken from \cite{Schruba11}.}
\label{fig2}
\end{center}
\end{figure}

The pixel study by Bigiel {et~al.} was however only sensitive to H$_2$ surface densities that are comparable to the saturation value of HI surface densities and therefore could only confirm the tight relationship between SFR and H$_2$ in ISM regions that are predominantly H$_2$. To determine the role of the H$_2$ gas phase for the star formation process in regimes that are predominantly HI required more sensitive H$_2$ measurements. This has been achieved by applying a novel stacking technique that first removes the Doppler shift from the observed CO spectra  due to galaxy rotation which then allows for coherent stacking of the shifted spectra inside rings of galactocentric radius (or any other selection parameter). This analysis resulted in sensitive measurements of $\Sigma_{\rm H2}$ down to $\sim$1 M$_\odot$~pc$^{-2}$ and revealed the roughly unchanged extension of the linear $\Sigma_{\rm SFR} - \Sigma_{\rm H2}$ relation previously observed in H$_2$-dominated regimes far into regions where the ISM is mostly HI (Fig.\,\ref{fig2} left panel; \cite{Schruba11}). This analysis also linked the turnover in the $\Sigma_{\rm SFR} - \Sigma_{\rm Gas}$ relation (Fig.\,\ref{fig1} right) -- previously interpreted as a ``threshold'' for star formation -- to a continuous change in the dominant ISM gas phase. This becomes most clear in the right panel of Fig.\,\ref{fig2} which shows the $\Sigma_{\rm SFR}/\Sigma_{\rm Gas}$ ratio of function of the $\Sigma_{\rm H2}/\Sigma_{\rm HI}$ ratio where the dashed lines indicate the expected trend if SFR scales linearly with H$_2$. The SFR per unit total gas mass thus depends largely on how molecular the ISM is -- which itself is controlled by a complex balance of H$_2$ formation and shielding from UV radiation that fundamentally depends on the dust-to-gas ratio or equivalently the metallicity. The determination of the exact dependencies is topic of active observational and theoretical research.

\section{Accounting for the Metallicity Dependence of the CO-H$_2$ Factor}

\begin{figure}[t]
\begin{center}
\begin{minipage}[b]{0.45\textwidth}
\includegraphics[width=1\textwidth]{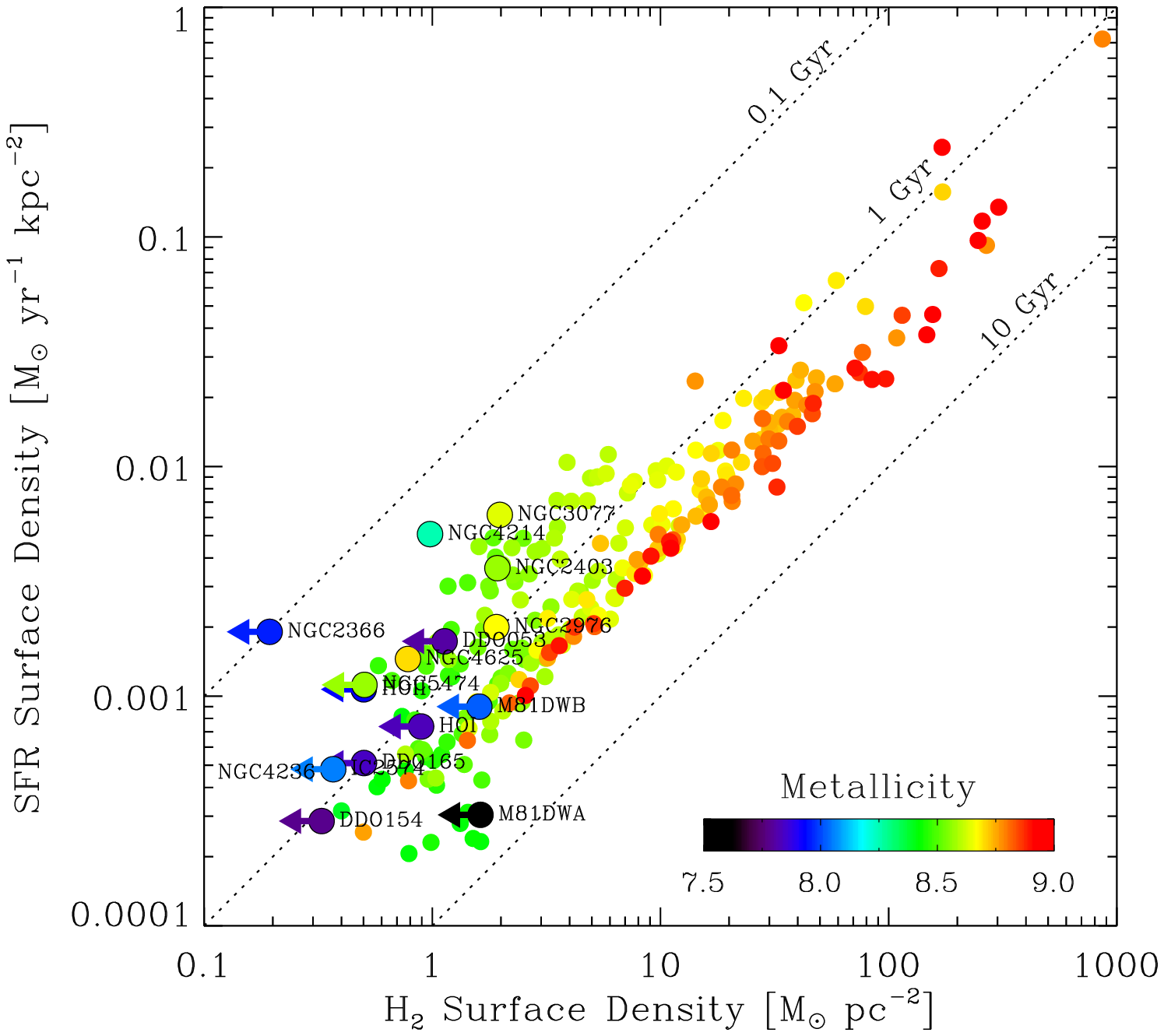}
\end{minipage}
\begin{minipage}[b]{0.45\textwidth}
\caption{The molecular star formation relation as a function of metallicity (color coded) when applying a constant CO-H$_2$ conversion factor. The spread in the relation is caused by (dwarf) galaxies of subsolar metallicity that are offset from the relation for massive solar metallicity spirals with depletion times of $\sim$2 Gyr. This offset stems by plotting observables (i.e., CO intensity) without accounting for the CO-H$_2$ conversion factor systematically changing with metallicity. Most of the scatter is removed if such a dependence is accounted for. Figure taken from \cite{Schruba12}.}
\vspace*{2 mm}
\end{minipage}
\label{fig3}
\end{center}
\end{figure}

Despite the fact that power law fits to the $\Sigma_{\rm SFR} - \Sigma_{\rm H2}$ data such as shown in the left panel of Fig.\,\ref{fig2} result in roughly linear relations, there is increasing scatter visible toward lower SFR and H$_2$ surface densities. This scatter is mostly caused by using CO emission to trace H$_2$ gas mass and the (inaccurate) use of a constant CO-H$_2$ conversion factor. Observations show that the CO/SFR ratio decreases steadily for regions of lower metallicities. The relative faintness of CO emission in low-metallicity environments persistently resulted in poor or failed CO detections in surveys of local dwarf galaxies (\cite{Young95}; \cite{Leroy05}). Employing the high sensitivity and large map coverage of the HERACLES data set and the stacking technique mentioned above, allowed us the first sensitive search for CO emission over the entire star-forming disk in local dwarf galaxies (\cite{Schruba12}). These measurements are shown in Fig.\,\ref{fig3} by the larger symbols with the radially averaged data for spiral galaxies shown as reference by the smaller symbols. Metallicity is shown by color coding and highlights a systematic trend in the CO/SFR ratio with metallicity.

\begin{figure}[t]
\begin{center}
\includegraphics[width=0.75\textwidth]{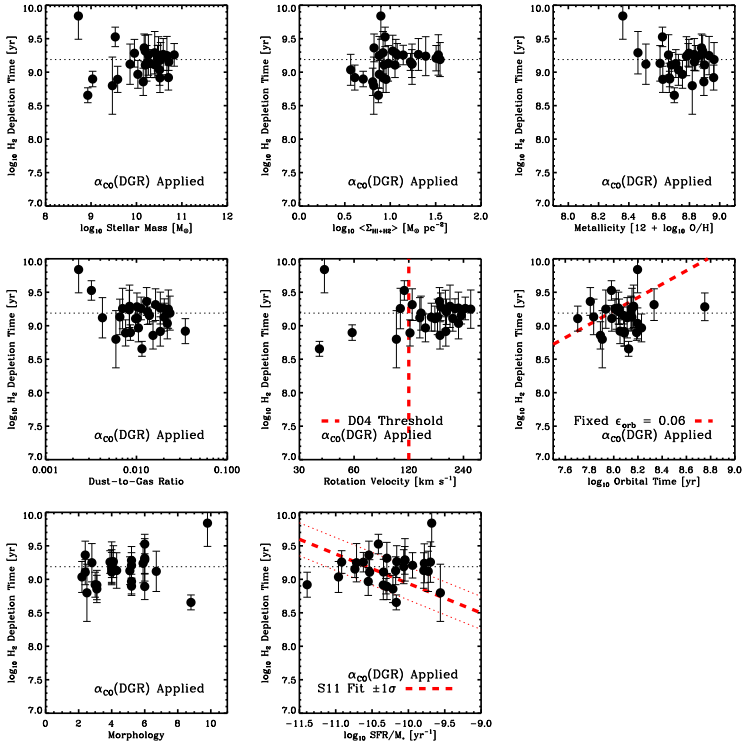}
\caption{The galaxy-averaged molecular star formation efficiency -- the ratio of SFR and H$_2$ mass -- as a function of various galaxy properties. A CO-H$_2$ conversion factor that scales with dust-to-gas ratio (or equivalently metallicity) has been applied following the model by \cite{Wolfire10} which is calibrated against the observed SFR/CO-DGR scaling. After accounting for systematic changes in the CO-H$_2$ factor, the SFR/H$_2$ ratio is found to be constant and independent of all here tested galaxy properties. Figure taken from \cite{Leroy12}.}
\label{fig4}
\end{center}
\end{figure}

The systematic trend with metallicity is however mostly an artifact of plotting (scaled) observables and vanishes once a metallicity dependent CO-H$_2$ conversion factor is applied. The exact metallicity (and other parameter) dependence of the CO-H$_2$ factor is still unclear today, however, approximate models are sufficient to remove most of the trends previously observed in the H$_2$/SFR ratio. Fig.\,\ref{fig4} highlights this by applying a CO-H$_2$ factor that scales with the dust-to-gas ratio as proposed by \cite{Wolfire10} and shows that the molecular gas depletion time is basically independent of a large set of galaxy properties that have been previously invoked to influence the star formation efficiency (\cite{Leroy12}). Thus most nearby spiral and dwarf galaxies seem to follow this ``normal mode of disk star formation'' with a molecular gas depletion time of $\sim$2 Gyr.

\section{Systematic Deviations from the Linear Star Formation Relation}

\begin{figure}[t]
\begin{center}
\begin{minipage}[b]{0.42\textwidth}
\includegraphics[width=1\textwidth]{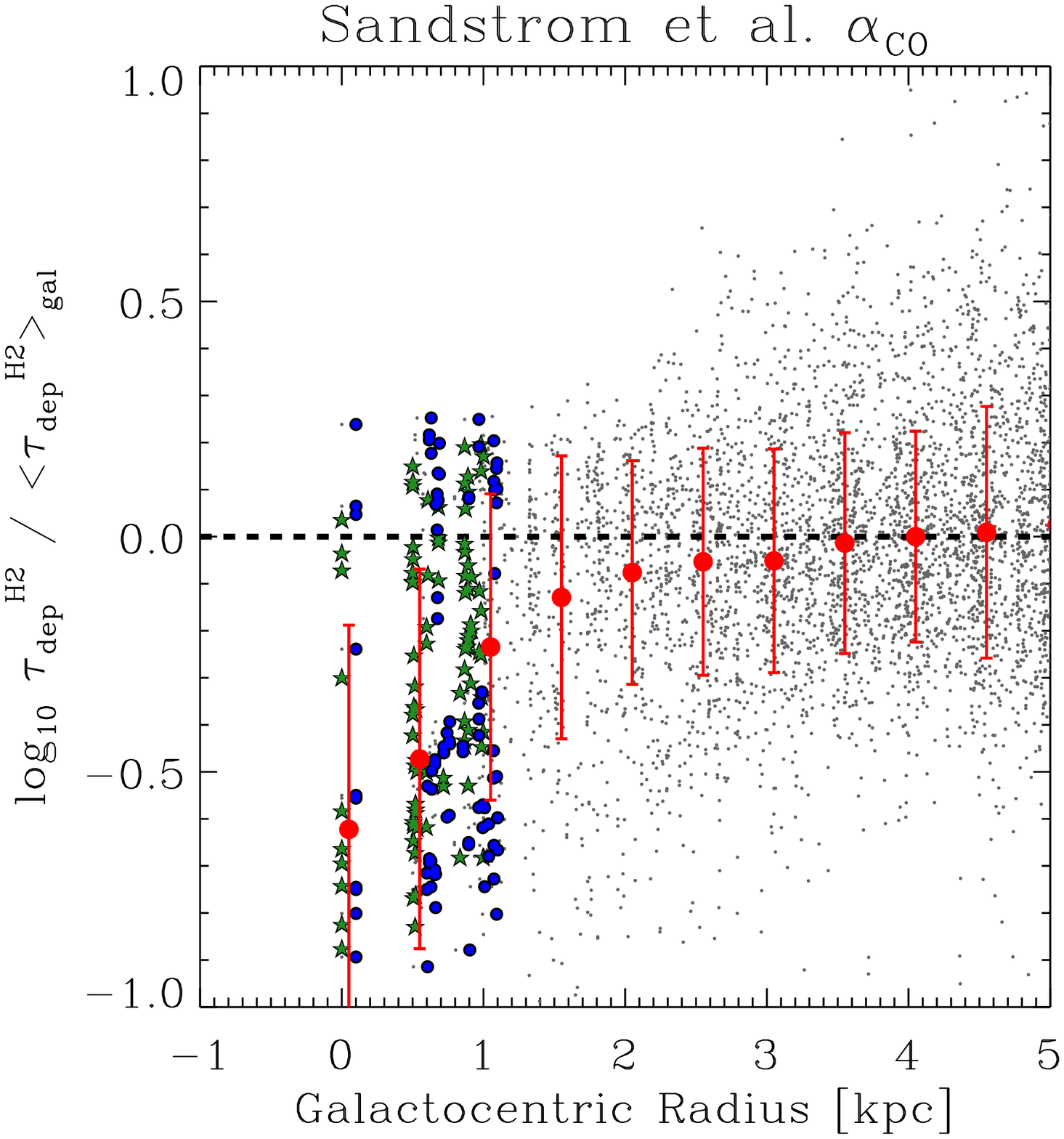}
\end{minipage}
\begin{minipage}[b]{0.4\textwidth}
\includegraphics[width=1\textwidth]{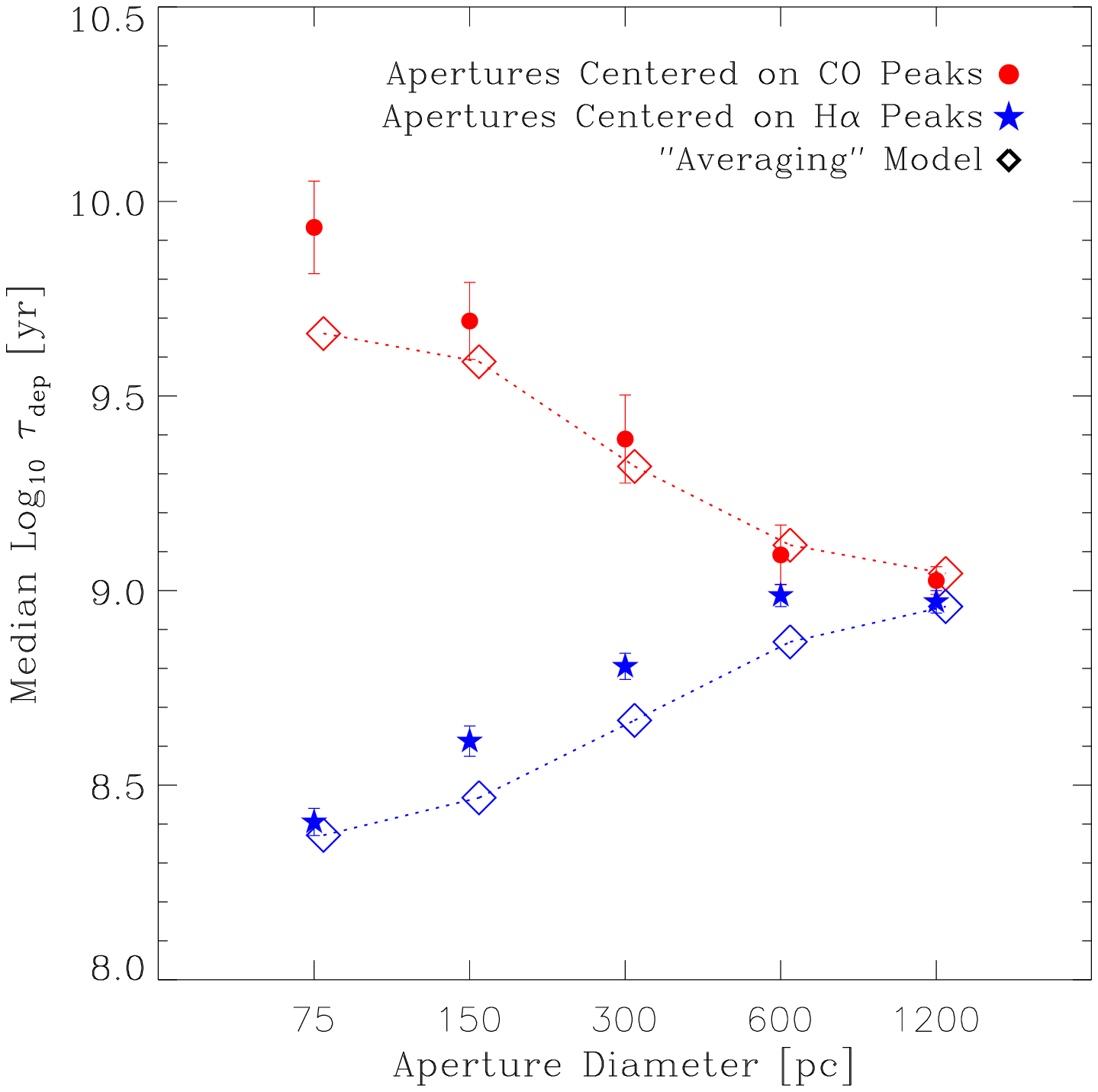}
\end{minipage}
\caption{Left: The normalized molecular star formation efficiency as function of galactocentric radius for 26 spiral galaxies using a CO-H$_2$ conversion factor that is locally calibrated against total gas masses inferred by dust modeling. Near the galaxy centers the SFR per unit molecular gas mass is often enhanced indicating a dependence of the star formation process on environment. Right: Deviations in the star formation relation also appear at small spatial scales. For a scale dependent aperture study in M33, the median ratio of CO/H$\alpha$ -- a proxy of the ratio of H$_2$/SFR -- varies by more than a magnitude between apertures centered on HII regions and CO peaks. The tight SFR-H$_2$ relation emerges only for larger apertures that include and average over several star-forming regions. Figures taken from \cite{Leroy12} and \cite{Schruba10}.}
\label{fig5}
\end{center}
\end{figure}

There are however several regimes where SFR and H$_2$ do not follow the tight relationship observed for spiral galaxy disks. One such deviation is frequently observed in galaxy centers which show an excess in the SFR/CO ratio. It is not trivial to access if the same deviation also translates to the SFR/H$_2$ ratio as the CO-H$_2$ conversion factor most likely changes in the central regions of galaxies because of higher gas temperatures and higher gas velocity dispersions (\cite{Narayanan11}; \cite{Narayanan12}). One method of obtaining a CO-independent determination of the molecular gas mass and thus constraining the CO-H$_2$ factor uses infrared continuum emission and dust modeling. \cite{Sandstrom12} applied this method to a sample of 26 spiral galaxies and \cite{Leroy12} used these results to show that an excess of SFR per unit H$_2$ in the centers of galaxies is a common feature (left panel of Fig.\,\ref{fig5}). This excess of star formation is likely caused by both higher dense gas fractions as well as higher intrinsic star formation efficiencies and thus resembles the ``starburst mode of star formation'' observed in local LIRGs and ULIRGs (e.g., \cite{GarciaBurillo12}). The reason what processes drive the enhanced dense gas fractions and star formation efficiencies are currently still unclear and topic of active research.

Another regime where the tight SFR-H$_2$ relation apparently breaks down is observed when looking at the relationship at high spatial resolution. The right panel of Fig.\,\ref{fig5} shows this by comparing the H$_2$/SFR ratio for apertures of various sizes that are centered either on either H$\alpha$-peaks of CO-peaks in M33 (\cite{Schruba10}). For individual star-forming regions, the H$_2$/SFR ratio systematically varies by more than an order of magnitude depending on which type of peak the aperture is centered on. The overall tight relationship is only recovered once apertures contain and average over several star-forming regions. On the other hand, the scale-dependence of the scatter in the star formation relation also contains valuable information on the time evolution of individual star-forming regions, that are thought to evolve from being CO-bright but quiescent when they form, evolve to have bright CO and H$\alpha$ emission coexisting for some time, and finally end as HII regions that removes the molecular gas (\cite{Kawamura09}). Rigorous attempts to extract this information from the rich data sets on the star formation relation are currently still sparse and will hopefully be extended in the near future (see \cite{Feldmann10} and \cite{Feldmann11} for a first study).

\section{Summary: The Composite Star Formation Relation}

\begin{figure}[t]
\begin{center}
\includegraphics[width=0.75\textwidth]{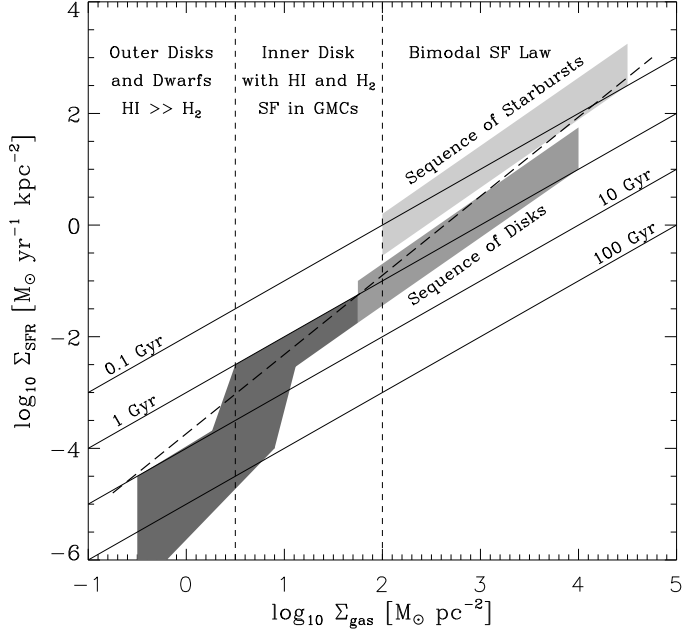} 
\caption{Sketch of the resolved star formation relation of different parts and types of galaxies. Three different regimes can be defined: The ``outer disks \& dwarfs'' regime (left) in which the decreasing availability of (atomic) gas limits the SFR with depletion times exceeding the Hubble time. The ``normal \& massive disks'' regime (middle) in which SFR correlates closely with H$_2$ with depletion times of $\sim$2 Gyr. The relation is linear as long as star formation proceeds in isolated molecular clouds and becomes super-linear once these merge to a continuous fluid. An ``inflated/bimodal'' trend (right) emerges at high gas surface densities in which the SFR per unit molecular gas reflects the varying dense gas fractions in quiescent disks and merging starbursts.}
\label{fig6}
\end{center}
\end{figure}

New detailed observations in the last few years have shown that the SFR is not a simple function of the total gas mass but has a complex behavior depending on the dominant gas phase. Fig.\,\ref{fig6} tries to make a sketch of these different regimes, though in a much oversimplified format. In regions of low gas surface densities, $\Sigma_{\rm Gas} \lesssim 1-5$ M$_\odot$~pc$^{-2}$, as found in the outer disks of spiral galaxies and in dwarf galaxies, $\Sigma_{\rm SFR}$ scales with $\Sigma_{\rm Gas}$ with gas consumption times much larger than the Hubble time (e.g., \cite{Bigiel10}). Star formation is limited by the availability of its raw material but the large scatter in the relation indicates that the ability of the ISM to form overdense, gravitational-unstable regions strong depends on environment. Towards smaller galactocentric radii and higher gas surface densities, $\Sigma_{\rm Gas} \sim 5-10$ M$_\odot$~pc$^{-2}$, the SFR increases steeply and is no longer predictable from the total gas mass. In this range the ISM consists both of atomic and molecular gas and its phase balance is a sensitive function of gas density, metallicity, and potentially other parameters. Here our sensitive stacking analysis finds $\Sigma_{\rm SFR} \propto \Sigma_{\rm H2}$ which is independent of the local atomic gas mass suggesting that the SFR per total gas mass is solely controled by the formation of a molecular gas phase. In the inner disks of spirals where $\Sigma_{\rm Gas} \gtrsim 10$ M$_\odot$~pc$^{-2}$ and the ISM is predominantly molecular, the SFR scales linearly with the available gas mass showing no significant dependence of galaxy type and/or global galaxy properties. At surface densities of $\Sigma_{\rm Gas} \gtrsim 100$ M$_\odot$~pc$^{-2}$ -- exceeding the surface density of GMCs -- the relation begins to steepen again but the uniform trend is broken. Instead the star formation relations shows a inflated/bimodal form and $\Sigma_{\rm SFR}$ can vary by factor $\sim 10$ at fixed gas surface density. First indications suggest that the enhancement in the star formation efficiency may be caused by a combination of higher dense gas fractions and higher star formation efficiencies. This enhanced ``starburst mode of star formation'' is likely a transient phase during galaxy evolution (most likely induced by mergers) at which the distribution of gas volume densities is shifted toward significantly higher values and thus enables more rapid star formation than normal in spirals.


\acknowledgments
The author thanks the SOC and LOC for organizing this stimulating and productive meeting. Furthermore, the author thanks Adam Leroy for providing the SFE-Environment plot and the SFE-Radius plot before publication. A.S. gratefully acknowledges financial support from the IAU.



\end{document}